\documentclass[letter,longauth]{aa} 

\usepackage{graphicx}
\usepackage{txfonts}
\usepackage[utf8]{inputenc}
\usepackage{array}
\usepackage{float}
\usepackage{multirow}
\usepackage[]{natbib}
\usepackage{amsmath}
\usepackage{epstopdf}
\usepackage[colorlinks=true,linkcolor=blue,citecolor=blue]{hyperref}
\newcommand{\rev}[1]{#1} 

\begin{document} 

\title{A dusty benchmark brown dwarf near the ice line of HD\,72946\thanks{Based on observations collected at the European Organisation for Astronomical Research in the Southern Hemisphere under ESO programme 0102.C-0781.}}

  \author{A.-L. Maire\inst{1,\thanks{F.R.S.-FNRS Postdoctoral Researcher.}},
  J.-L. Baudino\inst{2}, S. Desidera\inst{3}, S. Messina\inst{4}, W. Brandner\inst{5}, N. Godoy\inst{6,7}, F. Cantalloube\inst{5}, R. Galicher\inst{8}, M. Bonnefoy\inst{9}, J. Hagelberg\inst{10}, J. Olofsson\inst{6,7}, O. Absil\inst{1,\thanks{F.R.S.-FNRS Research Associate.}}, G. Chauvin\inst{9,11}, T. Henning\inst{5}, and M. Langlois\inst{12}
         }

   \institute{STAR Institute, Universit\'e de Li\`ege, All\'ee du Six Ao\^ut 19c, B-4000 Li\`ege, Belgium \\
          \email{almaire@uliege.be}
          \and
          Department of Physics, University of Oxford, Oxford, UK
          \and
          INAF--Osservatorio Astronomico di Padova, Vicolo dell'Osservatorio 5, I-35122 Padova, Italy
                 \and
          INAF Catania Astrophysical Observatory, via S. Sofia 78, 95123 Catania, Italy
          \and
          Max-Planck-Institut f\"ur Astronomie, K\"onigstuhl 17, D-69117 Heidelberg, Germany
          \and
                 Instituto de F\'isica y Astronom\'ia, Facultad de Ciencias, Univ. de Valpara\'iso, Av. Gran Breta\~{n}a 1111, Playa Ancha, Valpara\'iso, Chile
                 \and
                 N\'ucleo Milenio Formaci\'on Planetaria - NPF, Univ. de Valpara\'iso, Av. Gran Breta\~{n}a 1111, Playa Ancha, Valpara\'iso, Chile
          \and
          LESIA, Observatoire de Paris, PSL Research University, CNRS, Sorbonne Universit\'es, UPMC Univ. Paris 06, Univ. Paris Diderot, Sorbonne Paris Cit\'e, 5 place Jules Janssen, F-92195 Meudon, France
          \and
          Univ. Grenoble Alpes, CNRS, IPAG, F-38000 Grenoble, France
          \and
          Geneva Observatory, University of Geneva, Chemin des Maillettes 51, 1290 Versoix, Switzerland
          \and
          Unidad Mixta Internacional Franco-Chilena de Astronom\'ia CNRS/INSU UMI 3386 and Departamento de Astronom\'ia, Universidad de Chile, Casilla 36-D, Santiago, Chile
                 \and
                 CRAL, UMR 5574, CNRS/ENS-Lyon/Universit\'e Lyon 1, 9 av. Ch. Andr\'e, F-69561 Saint-Genis-Laval, France
            }

   \date{Received 18 November 2019 / Accepted 4 December 2019}

 
  \abstract
   {HD\,72946 is a bright and nearby solar-type star hosting a low-mass companion at long period ($P$\,$\sim$\,16~yr) detected with the radial velocity (RV) method. The companion has a minimum mass of 60.4$\pm$2.2~$M_{\rm{J}}$ and might be a brown dwarf. Its expected semi-major axis of $\sim$243~mas makes it a suitable target for further characterization with high-contrast imaging, in particular to measure its inclination, mass, and spectrum and thus definitely establish its substellar nature.}
   {We aim to further characterize the orbit, atmosphere, and physical nature of HD\,72946B.}
   {We \rev{present} high-contrast imaging data in the near-infrared with the Spectro-Polarimetric High-contrast Exoplanet REsearch (SPHERE) instrument. We also \rev{use} proper motion measurements of the star from \textsc{Hipparcos} and \textit{Gaia}.}
  {The SPHERE data reveal a point source with a contrast of $\sim$9~mag at a projected separation of $\sim$235~mas. No other point sources are detected in the field of view. By jointly fitting the RV, imaging, and proper motion data, we constrain all the orbital parameters of HD~72946B and assess a dynamical mass of 72.4$\pm$1.6~$M_{\rm{J}}$ \rev{and a semi-major axis of 6.45$^{+0.08}_{-0.07}$~au}. Empirical comparison of \rev{its SPHERE} spectrum to template dwarfs indicates a spectral type of L5.0$\pm$1.5. The $J$-$H3$ color is close to the expectations of the DUSTY models and suggests a cloudy atmosphere. Comparison with atmospheric models of the spectrophotometry suggests an effective temperature of $\sim$1700~K. The bolometric luminosity (log($L$/$L_{\sun}$)\,=\,-4.11$\pm$0.10~dex) and dynamical mass of HD\,72946B are more compatible with evolutionary models for an age range of $\sim$0.9--3~Gyr. The formation mechanism of \rev{the companion} is currently unclear as \rev{the object appears slightly away from the bulk of model predictions}. HD\,72946B is currently the closest benchmark brown dwarf companion to a solar-type star with imaging, RV, and proper motion measurements.}
   {}


\authorrunning{A.-L. Maire et al.}
\titlerunning{A dusty benchmark brown dwarf near the ice line of HD\,72946}

   \maketitle

\section{Introduction}

\begin{table*}[t]
\caption{\rev{Relative photometry and astrometry of HD\,72946B}.}
\label{tab:photoastrometry}
\begin{center}
\begin{tabular}{l c c c c c c c c}
\hline\hline
Filter & $\lambda_0$ & $\Delta\lambda$ & $\Delta$mag & Abs. mag. & Flux & Separation & PA \\
 & ($\muup$m) & ($\muup$m) & (mag) & (mag) & ($\times$10$^{-15}$ W\,m$^{-2}$\,$\muup$m$^{-1}$) & (mas) & ($^{\circ}$) \\
\hline
$H2$ & 1.593 & 0.052 & 8.97$\pm$0.07 & 12.51$\pm$0.07 & 3.027$\pm$0.188 & 235.7$\pm$2.0 & 33.65$\pm$0.31 \\
$H3$ & 1.667 & 0.054 & 8.81$\pm$0.07 & 12.35$\pm$0.07 & 3.236$\pm$0.204 & 235.6$\pm$2.0 & 33.68$\pm$0.31 \\
\hline
\end{tabular}
\end{center}
\tablefoot{The photometric error bars were derived assuming an error budget including the measurement uncertainties (image post-processing) and systematic uncertainties (temporal variability of the flux calibration and of the science sequence). \\
}
\end{table*}

Dynamical mass measurements of brown dwarfs are a powerful test of their formation and evolution models. Most studies \rev{exploit} brown dwarf binaries \citep[e.g.,][]{Konopacky2010, Dupuy2017, Dieterich2018}, which have likely formed by fragmentation of a collapsing cloud \citep[e.g.,][]{Bate2009}. However, it is still unclear whether brown dwarfs found at close-in separations to stars form like stellar binaries or by disk gravitational instabilities \citep{Boss1997}. In the past years, a few radial velocity (RV) surveys started to target stars with slow drifts to constrain the orbit and minimum mass of the suspected long-period companions \citep[e.g.,][]{Bouchy2016, Sahlmann2011, Feroz2011}. These surveys have shown a paucity of brown dwarf companions within 5~au from the host stars with respect to planetary and stellar companions \citep[the so-called ``brown dwarf desert'', see e.g.,][]{Grether2006, Sahlmann2011, Ma2014}. Nevertheless, \citet{Ma2014} \rev{found} that their occurrence increases at larger separations when brown dwarf detections from various techniques are combined.

Using the ELODIE and SOPHIE instruments, \citet{Bouchy2016} \rev{reported} a potential brown dwarf companion to the G5V star HD\,72946, located at 25.87$\pm$0.08~pc \citep{GaiaCollaboration2016b, GaiaCollaboration2018}. The RV data cover a full orbit of HD\,72946B, which \rev{allowed} the authors to place good constraints on its orbit (period $P$\,=\,15.93$^{+0.15}_{-0.13}$~yr, eccentricity $e$\,=\,0.495$\pm$0.006, and periastron $T_0$ \rev{[HJD]} \,=\,2455958$\pm$10). They \rev{derived} a minimum dynamical mass of 60.4$\pm$2.2~$M_{\rm{J}}$ and an upper mass limit of 0.2~$M_\odot$ from the analysis of the cross-correlation function of the star.

We present in this paper the confirmation and characterization of \rev{the brown dwarf} companion to HD\,72946 with the Spectro-Polarimetric High-contrast Exoplanet REsearch (SPHERE) instrument and \textsc{Hipparcos}-\textit{Gaia} proper motion measurements. We present an updated analysis of the properties of the host star in Sect.~\ref{sec:star_ppties} and the SPHERE imaging observations in Sect.~\ref{sec:data}. We perform a joint orbital fit of the imaging, RV, and astrometric data and derive a dynamical mass for HD\,72946B in Sect.~\ref{sec:orbit}. Section~\ref{sec:sed} discusses the spectral properties of the companion. Finally, we compare the physical and spectral properties of HD\,72946B to model predictions in Sect.~\ref{sec:compa_models}. 

\section{Properties of the host star}
\label{sec:star_ppties}

\citet{Bouchy2016} \rev{inferred} from spectroscopic observations an effective temperature $T_{\rm{eff}}$ = 5686$\pm$ 40~K, a surface gravity log\,$g$ = 4.50$\pm$0.06~dex, and a metallicity [Fe/H] = 0.11$\pm$0.03~dex. The supersolar metallicity has been confirmed by other studies \citep[][0.15$\pm$0.06, 0.16$\pm$0.04, and 0.12~dex, respectively]{AguileraGomez2018, Luck2006, Casagrande2011}. 

We \rev{derived} the stellar age and mass from isochrones using the PARAM web interface\footnote{\url{http://stev.oapd.inaf.it/cgi-bin/param_1.3}} \citep{daSilva2006}. We adopted the spectroscopic $T_{\rm{eff}}$  and [Fe/H] in \citet{Bouchy2016} with enlarged uncertainties to account for systematic errors. \rev{We also adopted the \textit{Gaia} parallax and the $V$-band magnitude from \textsc{Hipparcos} (7.08$\pm$0.02~mag)}. This results in an age 1.712$\pm$1.684~Gyr, a mass 0.986$\pm$0.027~$M_{\sun}$, and a radius 0.908$\pm$0.018~$R_{\sun}$. Tighter constraints on the age from isochrones can be derived from the F8 comoving companion HD\,72945 (1.6$\pm$1.0~Gyr, Appendix~\ref{sec:multiplicity}).

Lithium data \citep[A(Li) = 1.41, 1.23, 1.22$\pm$0.15~dex,][]{Luck2017, Luck2006, Ramirez2012} indicate an age older than that of the Hyades and similar to the open cluster NGC\,752 \citep{Sestito2004}. The stellar kinematics suggest an age younger than the Sun, the UVW velocities being at the boundary of the kinematic space of young stars in \citet{Montes2001a}. Comparisons with stars with similar kinematics in \citet{Casagrande2011} indicated that it is unlikely that the star is older than $\sim$3~Gyr and much younger than 0.5~Gyr.

We searched for archival photometric data to derive an age with gyrochronology, but we did not find suitable data (sampling, accuracy, blending with HD\,72945, and/or calibration issues). Using the relations in \citet{Mamajek2008} and an averaged measured chromospheric activity of -4.60~dex \citep[individual values -4.54, -4.74$\pm$0.05, -4.66, and -4.47~dex,][]{RochaPinto2004, Bouchy2016, Gray2003, BoroSaikia2018}, we derive a rotation period of $\sim$15~d, which implies a gyrochronological age of $\sim$1~Gyr. This is in between the loci of the Hyades (625--700~Myr) and NGC\,752 (2000~Myr). The star has X-ray data from ROSAT, but is blended with HD\,72945. However, X-ray activity is expected to correlate with chromospheric activity, so that it does not provide a fully independent age estimate. Assuming our derived stellar radius and an averaged measured projected rotational activity of 4.14~km\,s$^{-1}$ \citep[individual values 3.23, 3.9$\pm$1, and 5.3~km\,s$^{-1}$,][]{MartinezArnaiz2010, Bouchy2016, Luck2017}, we derive an upper limit for the rotation period of 12~d, which implies a gyrochronological age younger than 1~Gyr considering a $B-V$ color of 0.71~mag. Considering the large uncertainties in $v$sin\,$i_{\star}$, the upper limit for the gyrochronological age could be as old as 1.5~Gyr. This means that our various age estimates agree overall. In the following, we \rev{choose} to adopt an age range of 0.8--3~Gyr, with a most probable value of 1--2~Gyr.

\section{Observations and data analysis}
\label{sec:data}

\begin{figure}[t]
\centering
\includegraphics[width=.22\textwidth]{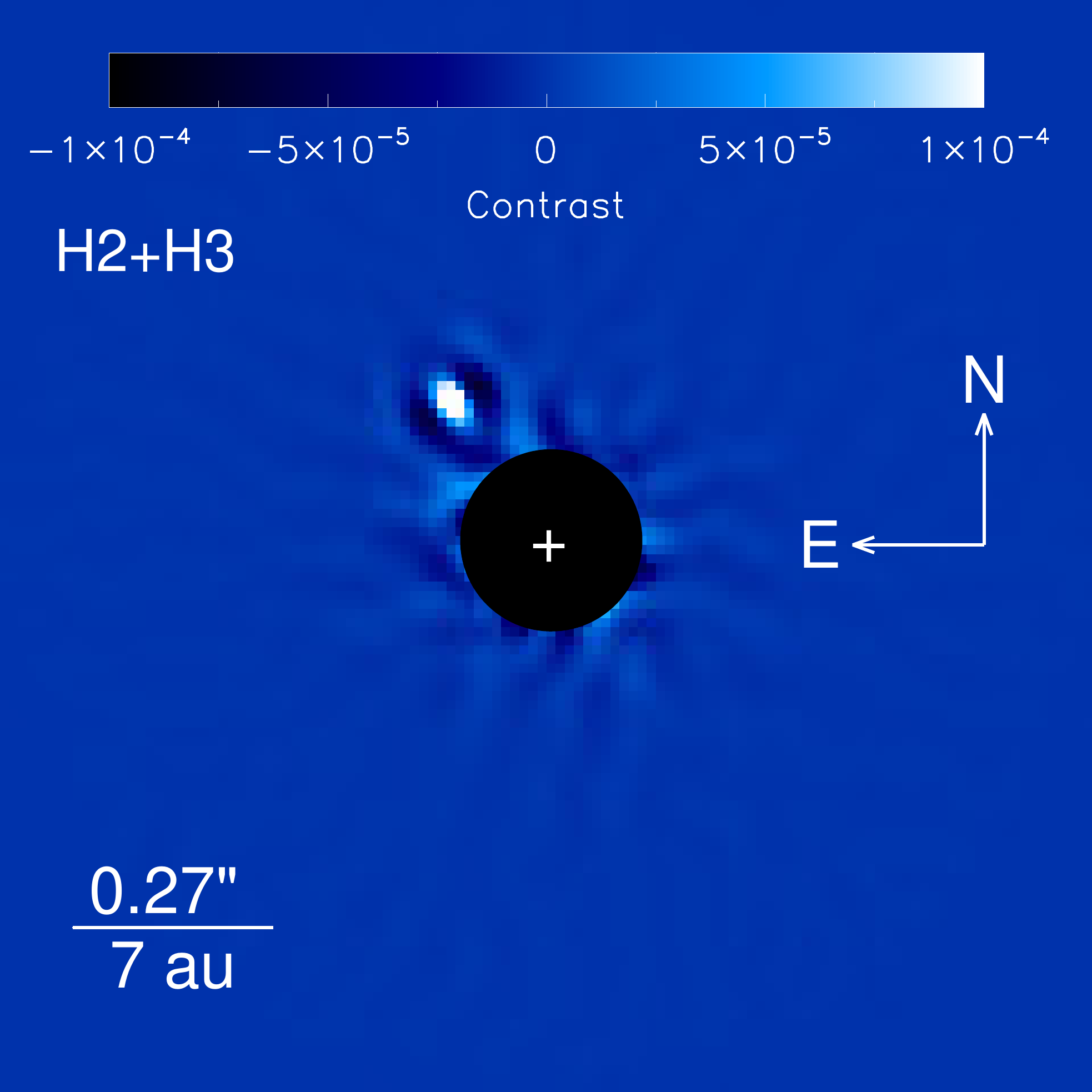}
\includegraphics[width=.22\textwidth]{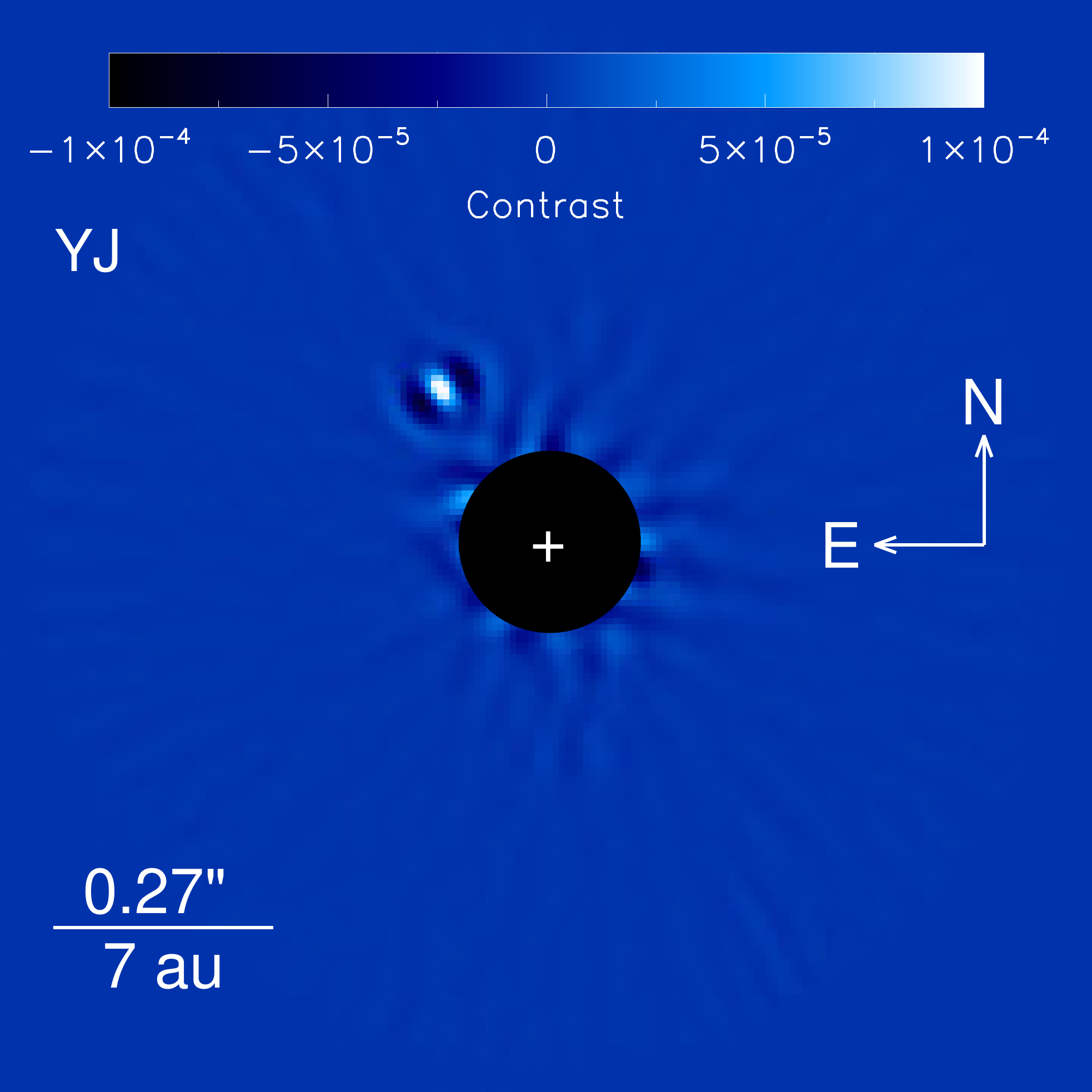}
\caption{\rev{SPHERE contrast images of HD\,72946. The central regions of the images are numerically masked out to hide bright stellar residuals. The white crosses indicate the location of the primary star.}}
\label{fig:images}
\end{figure}

\begin{figure*}[t]
\centering
$\vcenter{\hbox{\includegraphics[trim = 6mm 0mm 19mm 10mm, clip, width=.35\textwidth]{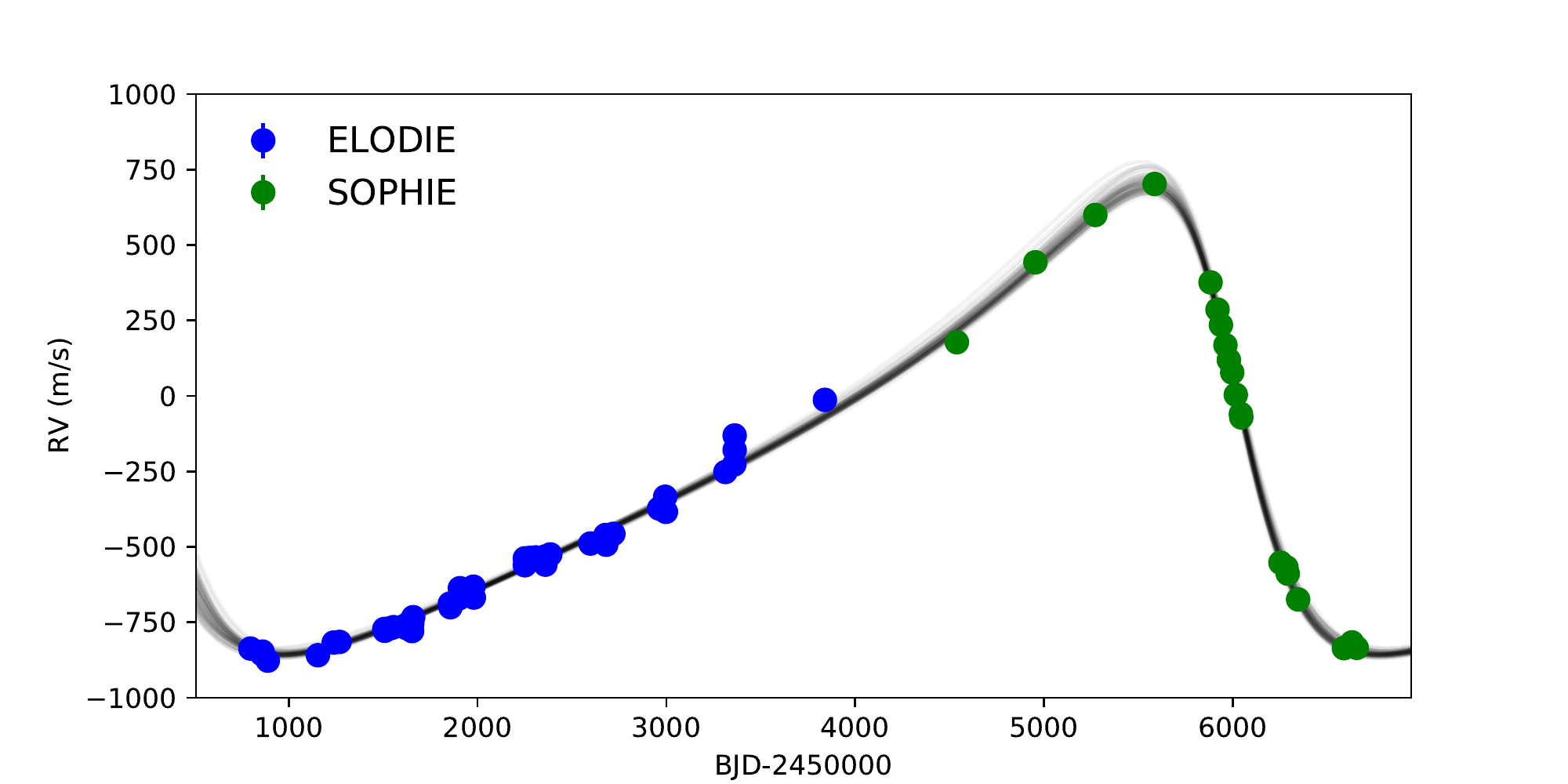}}}$
\hspace*{.02in}
$\vcenter{\hbox{\includegraphics[trim = 0mm 6mm 10mm 19mm, clip, width=.31\textwidth]{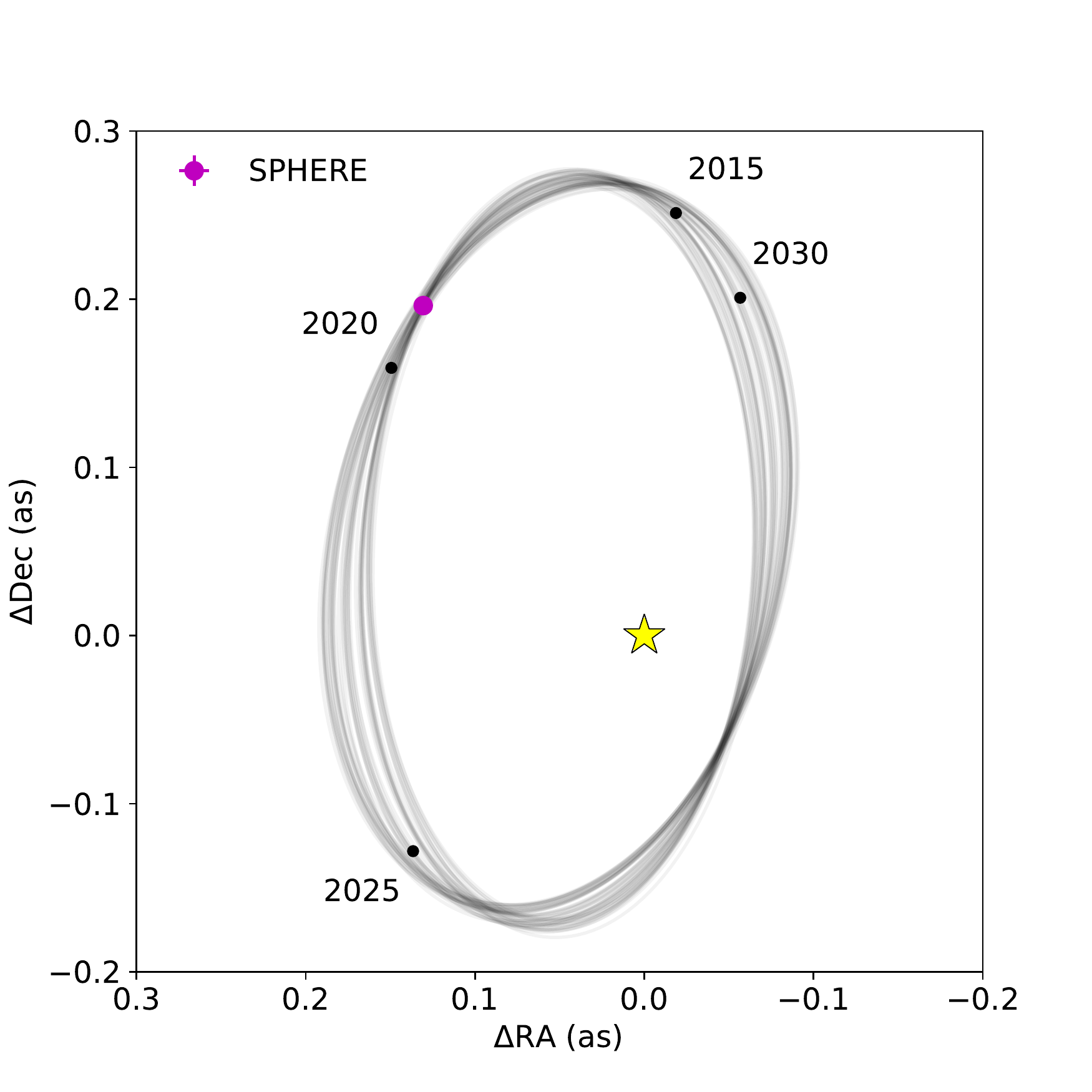}}}$
\hspace*{.02in}
$\vcenter{\hbox{\includegraphics[trim = 2mm 4mm 10mm 17mm, clip, width=.31\textwidth]{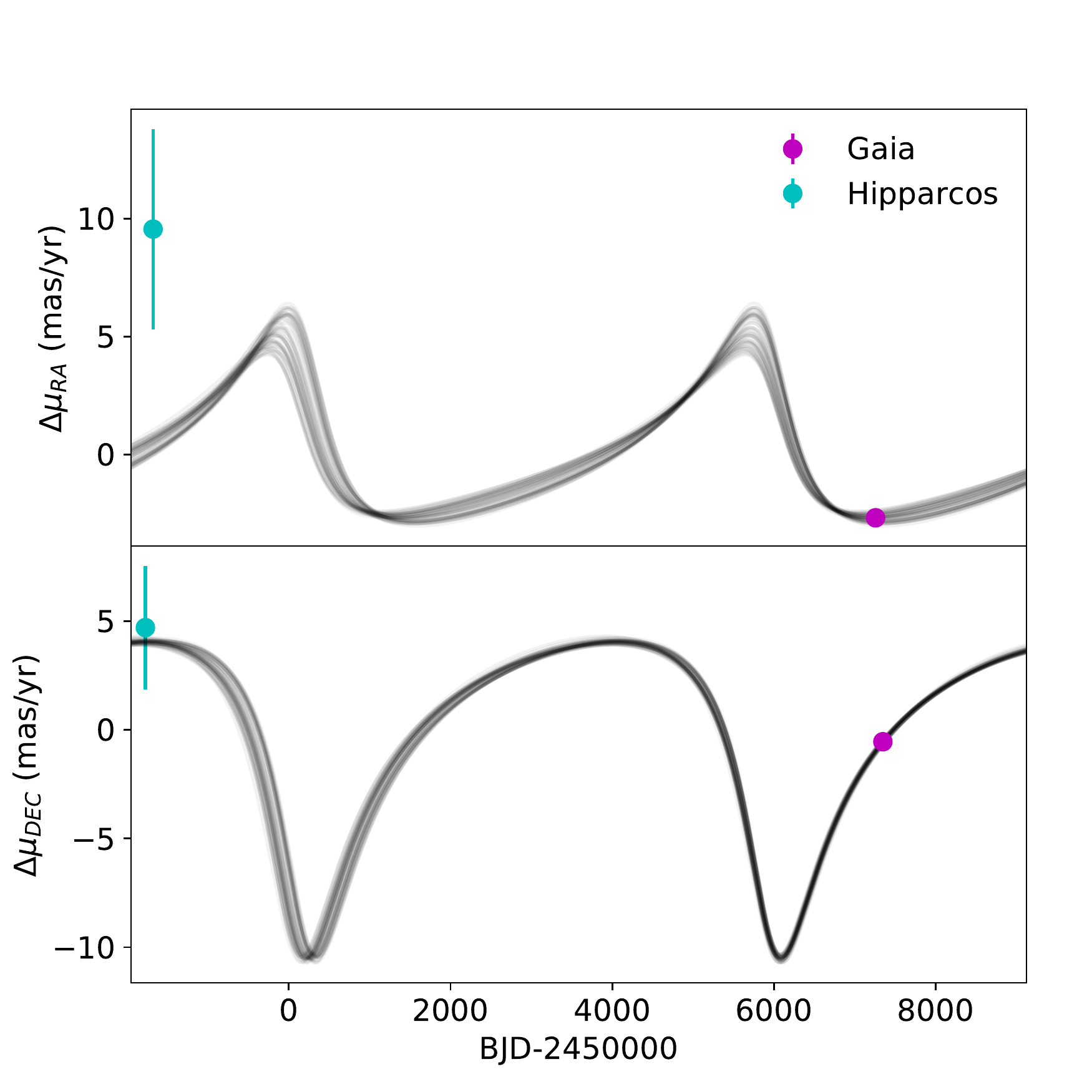}}}$
\caption{Sample of 50 model orbits fitted on the HD\,72946B data (colored points) from RV (\textit{left}), imaging (\textit{middle}), and astrometry (\textit{right}). In the \textit{middle} panel, the yellow star marks the location of the \rev{primary star,} and the black dots show the median predicted position for given epochs.}
\label{fig:modelorbits}
\end{figure*}

We observed HD\,72946 on 2019 March 21 UT with the standard IRDIFS mode of SPHERE \citep{Beuzit2019}, which allows for simultaneous near-IR observations with IRDIS with the $H23$ filter pair \citep{Dohlen2008a, Vigan2010} and the integral field spectrograph IFS in the $YJ$ bands \citep{Claudi2008}. The seeing and coherence time measured by the differential image motion monitor at 0.5~$\muup$m were 0.5--0.7$''$ and 6-8~ms, respectively. The detector integration time was set to 16~s, and 128 frames were recorded, amounting to a field rotation of 15.5$^{\circ}$.

An apodized pupil Lyot coronagraph \citep{Carbillet2011, Martinez2009} was used. We acquired data before and after the sequence to calibrate the flux of the images and the location of the star behind the coronagraph \citep{Langlois2013}. Night-time sky background frames were taken and additional daytime calibration performed following the standard procedure at ESO.

The data were reduced with the SPHERE Data Reduction and Handling software \citep[v0.15.0,][]{Pavlov2008} and custom routines for IFS data adapted from \citet{Mesa2015} and \citet{Vigan2015}. This \rev{corrected} for the cosmetics and instrument distortion, registered the frames, and normalized their flux. For IFS, it also \rev{performed} the wavelength calibration and extracted the image cubes. Then, the data were analyzed with angular differential imaging \citep{Marois2006a} using three algorithms (Appendix~\ref{sec:compa_algos})\rev{: ANDROMEDA, TLOCI, and PCA}. Figure~\ref{fig:images} shows the ANDROMEDA images.

The photometry and astrometry were extracted using three algorithms, but we chose to retain the TLOCI values (Table~\ref{tab:photoastrometry}). The astrometry was calibrated following \citet{Maire2016b}, with pixel scales of 12.255$\pm$0.009~mas/pix ($H2$) and 12.251$\pm$0.009~mas/pix ($H3$) and a North correction angle of $-$1.75$\pm$0.08$^{\circ}$. The absolute magnitudes were computed using  the 2MASS values \citep{Cutri2003} for the stellar magnitudes.

\section{Orbital analysis}
\label{sec:orbit}

We retrieved the RV measurements in \citet{Bouchy2016} through the VizieR interface. With only one imaging data point, there is still an ambiguity in the inclination and longitude of the ascending node. To solve for this, we also searched for an astrometric signature of the companion in the \textsc{Hipparcos}-\textit{Gaia} catalog of accelerations \citep{Brandt2018, Brandt2019a}: pmra\_g\_hg = $-$2.837$\pm$0.140~mas\,yr$^{-1}$ and pmdec\_g\_hg = $-$0.515$\pm$0.082~mas\,yr$^{-1}$ for \textit{Gaia} \citep{GaiaCollaboration2018}, pmra\_h\_hg = 9.411$\pm$4.245~mas\,yr$^{-1}$ and pmdec\_h\_hg = 4.734$\pm$2.839~mas\,yr$^{-1}$ for \textsc{Hipparcos} \citep{Perryman1997, vanLeeuwen2007}. These values imply an astrometric detection at (20.3, 6.3)$\sigma$ with \textit{Gaia} and (2.1, 1.7)$\sigma$ with \textsc{Hipparcos}. We verified that the \textit{Gaia} DR2 record is well behaved, with a renormalized unit weight error below 1.4 \citep{Lindegren2018}.

We \rev{performed} a joint fit of the RV, imaging, and proper motion data with the parallel-tempered Markov chain Monte Carlo (MCMC) algorithm provided in the \texttt{emcee} package \citep{ForemanMackey2013}, which is based on the algorithm described by \citet{Earl2005}. Our implementation follows \citet{Brandt2019b} in the broad lines. We \rev{sampled} the parameter space of our 13-parameter model assuming 15 temperatures for the chains and 100 walkers. The first 8 parameters are the semi-major axis $a$, the eccentricity $e$ and argument of periastron passage $\omega$ (parameterized as $\sqrt{e}$\,cos\,$\omega$ and $\sqrt{e}$\,sin\,$\omega$), the inclination $i$, the longitude of the ascending node $\Omega$, the time at periastron passage $T_0$, the RV semi-amplitude of the star $\kappa_A$, and the systemic velocity $\gamma$. We present the results for $\Omega$ and $\omega$ as relative to the companion. To fit the imaging and proper motion data, we \rev{used} the equations in Appendix~A of \citet{Makarov2005}. 

The initial state of the sampler \rev{was} set assuming uniform priors in log\,$a$, $\sqrt{e}$\,cos\,$\omega$, $\sqrt{e}$\,sin\,$\omega$, $\Omega$, $T_0$, and $\kappa_A$, as well as a sin\,$i$ prior for $i$. The width of the priors \rev{were} selected from the results in \citet{Bouchy2016} and a fit to the RV and imaging data with a least-squares Monte Carlo approach \citep{Maire2015, Schlieder2016} to derive first ranges for $i$ and $\Omega$. We \rev{disentangled} the two ($i$,$\Omega$) solutions by comparing the predictions for the instantaneous stellar proper motions to the measurements.

\begin{table}[t]
\caption{Orbital parameters and dynamical mass of HD\,72946B.}
\label{tab:orbparams}
\begin{center}
\begin{tabular}{l c c c}
\hline\hline
Parameter & Unit & Median $\pm$ 1$\sigma$ & Best fit \\
\hline
\multicolumn{4}{c}{Fitted parameters} \\
\hline
Semi-major axis $a$ & mas & 249.1$^{+3.1}_{-3.0}$ & 250.4 \\[3pt]
$\sqrt{e}$\,cos\,$\omega$ & & 0.231$^{+0.019}_{-0.020}$ & 0.230 \\[3pt]
$\sqrt{e}$\,sin\,$\omega$ & & 0.662$^{+0.008}_{-0.009}$ & 0.663 \\[3pt]
Inclination $i$ & $^{\circ}$ & 59.3$^{+2.3}_{-2.0}$ & 59.6 \\[3pt]
PA of asc. node $\Omega$ & $^{\circ}$ & -12.0$^{+4.3}_{-3.9}$ & -11.8 \\[3pt]
Time periastron $T_0$ & \rev{BJD} & 2455956.7$^{+10.7}_{-10.1}$ & 2455955.1 \\[3pt]
RV semi-ampl. $\kappa_A$ & m\,s$^{-1}$ & 778.7$^{+10.5}_{-9.4}$ & 774.1 \\[3pt]
Syst. velocity $\gamma$ & m\,s$^{-1}$ & -203.2$^{+8.4}_{-8.6}$ & -207.4 \\[3pt]
Parallax $\pi$ & mas & 38.65$\pm$0.12 & 38.66 \\[1.5pt]
SMA primary $a_1$ & mas & 16.28$^{+0.29}_{-0.26}$ & 16.12 \\[3pt]
RV offset ZP$_{\rm{SOPHIE}}$ & m\,s$^{-1}$ & 90.6$^{+15.6}_{-16.9}$ & 96.2 \\[3pt]
RV jitter $\sigma_{\rm{ELODIE}}$ & m\,s$^{-1}$ & 24.4$^{+4.1}_{-3.3}$ & 22.9 \\[3pt]
RV jitter $\sigma_{\rm{SOPHIE}}$ & m\,s$^{-1}$ & 16.1$^{+5.0}_{-3.5}$ & 12.67 \\[3pt]
\hline
\multicolumn{4}{c}{Computed parameters} \\
\hline
$M_1$ & $M_{\odot}$ & 0.99$\pm$0.03 & 1.01 \\[1.5pt]
$M_2$ & $M_{\rm{J}}$ & 72.4$\pm$1.6 & 72.5 \\[1.5pt]
Mass ratio $M_2$/$M_1$ & & 0.070$\pm$0.002 & 0.069 \\[3pt]
Period $P$ & yr & 15.91$^{+0.16}_{-0.13}$ & 15.90 \\[3pt]
Semi-major axis $a$ & au & 6.45$^{+0.08}_{-0.07}$ & 6.48 \\[1.5pt]
Eccentricity $e$ & & 0.493$^{+0.007}_{-0.008}$ & 0.493 \\[3pt]
Arg. periastron $\omega$ & $^{\circ}$ & 250.7$\pm$1.7 & 250.9 \\[3pt]
\hline
\end{tabular}
\end{center}
\end{table}

The next two parameters in our model are the parallax and the semi-major axis of the orbit of the host star around the center of mass of the system. For the parallax, we \rev{drew} the initial guesses around the nominal value measured by \textit{Gaia} assuming a combination of a Gaussian distribution for the measurement uncertainties and a uniform distribution for potential systematics ($<$0.1~mas, \url{https://www.cosmos.esa.int/web/gaia/dr2}). We \rev{drew} the semi-major axis of the star around a guess value computed from its mass (0.99~$M_{\sun}$), the companion mass (0.07~$M_{\sun}$), and the total semi-major axis, assuming a uniform distribution with a half-width of 1.5~mas. The last free model parameters are one RV offset and two RV jitters, using the results in \citet{Bouchy2016} as first guesses.

We \rev{ran} the MCMC for 125\,000 iterations and \rev{verified} the convergence of the chains using the integrated autocorrelation time \citep{ForemanMackey2013, Goodman2010}. The posterior distributions in Appendix~\ref{sec:cornerplotorbit} \rev{were} obtained after thinning the chains by a factor 100 to mitigate the correlations and discarding the first 75\% of the chains as the burn-in phase. The median values with 1$\sigma$ uncertainties and the best-fit values of the parameters are given in Table~\ref{tab:orbparams}. The uncertainties in the parameters in common with \citet{Bouchy2016} are slightly larger or similar. A sample of model orbits is shown in Fig.~\ref{fig:modelorbits}.

\rev{We note that the proper motion anomaly measured by \textsc{Hipparcos} in RA is different by $\sim$2$\sigma$ from the orbital predictions, whereas the measurement in DEC is well reproduced within the uncertainties. The \textsc{Hipparcos} and \textit{Gaia} data affect the derived orbital parameters and dynamical mass within the uncertainties with respect to a fit that only uses the RV and imaging data, except for breaking the ambiguity in the inclination and longitude of ascending node.}

\section{Spectral analysis}
\label{sec:sed}

We \rev{used} the IRDIS dual-band photometry of the companion to compute the color-magnitude diagram in Appendix~\ref{sec:cmd} \citep[details from Appendix C of][]{Bonnefoy2018}. We note that HD\,72946B is located near mid-L template dwarfs and is close to HIP\,65426b \citep{Chauvin2017a}.

\begin{figure}[t]
\centering
\includegraphics[width=.42\textwidth, trim = 8mm 0mm 8mm 16mm, clip]{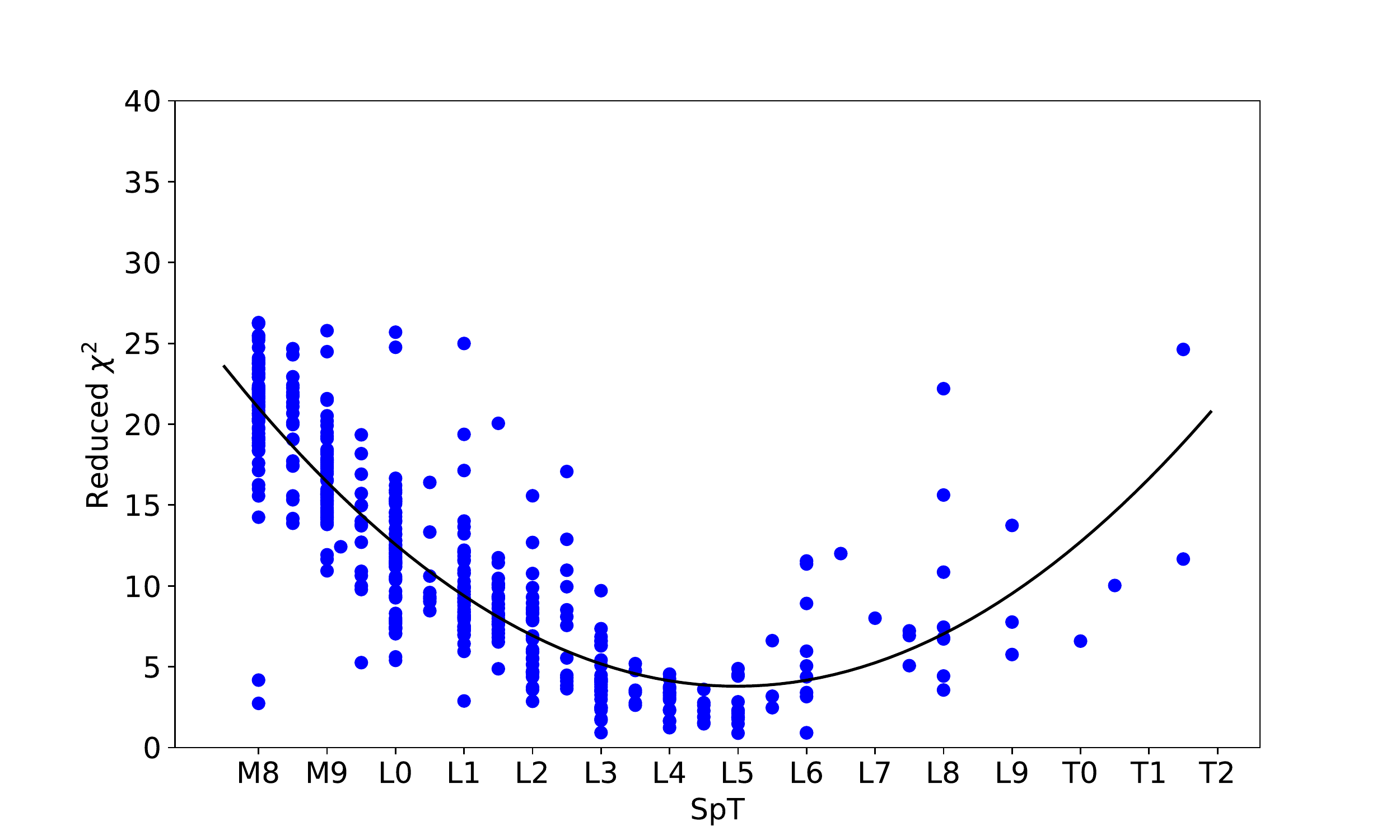}
\caption{Reduced $\chi^2$ as a function of the spectral type of the comparison of the IFS spectrum of HD\,72946B to SpeX template dwarfs.}
\label{fig:chi2_spt}
\end{figure}

We \rev{compared} the IFS spectrum to spectra of template dwarfs of the SpeX spectral library using the SPLAT toolkit \citep{2014ASInC..11....7B}. Figure~\ref{fig:chi2_spt} shows the reduced $\chi^2$ as a function of the spectral type. We \rev{include} the uncertainties of the template spectra in the $\chi^2$ computation. The best-fit object is the red L dwarf 2MASS J03552337+1133437 \citep{BardalezGagliuffi2014} (reduced $\chi^2$\,=\,0.89, assuming 38 degrees of freedom), which \rev{is} classified as L5$\gamma$ by \citet{Cruz2009}. From a parabolic fit to the $\chi^2$ values, we \rev{estimate} a spectral type of L5.0$\pm$1.5 considering all spectral types that satisfy $\chi^2$<$\chi^2_{\mathrm{min}}$+1.

To fit the \rev{spectrophotometry of HD\,72946B} with atmospheric models, we \rev{converted} \rev{the contrast measurements into physical fluxes using a model spectrum for the star} ($T_{\rm{eff}}$\,=\,5600~K, log\,$g$\,=\,4.5~dex, and [Fe/H]\,=\,0.0~dex) from the BT-NextGen library \citep{Allard2012} and the SPHERE filter transmission curves. \rev{The BT-NextGen spectrum is fit} to the stellar spectral energy distribution (SED) over the range 0.3--12~$\muup$m using the Virtual Observatory SED Analyzer \citep{Bayo2008}. The \rev{stellar SED} \rev{is} built using data from Tycho \citep{Hog2000}, 2MASS \citep{Cutri2003}, WISE \citep{2013yCat.2328....0C}, and IRAS \citep{Helou1988}, as well as Johnson photometry \citep{Mermilliod2006} and Str\"omgren photometry \citep{Paunzen2015}.

\begin{figure}[t]
\centering
\includegraphics[width=.42\textwidth,trim = 14mm 5mm 5mm 11mm, clip]{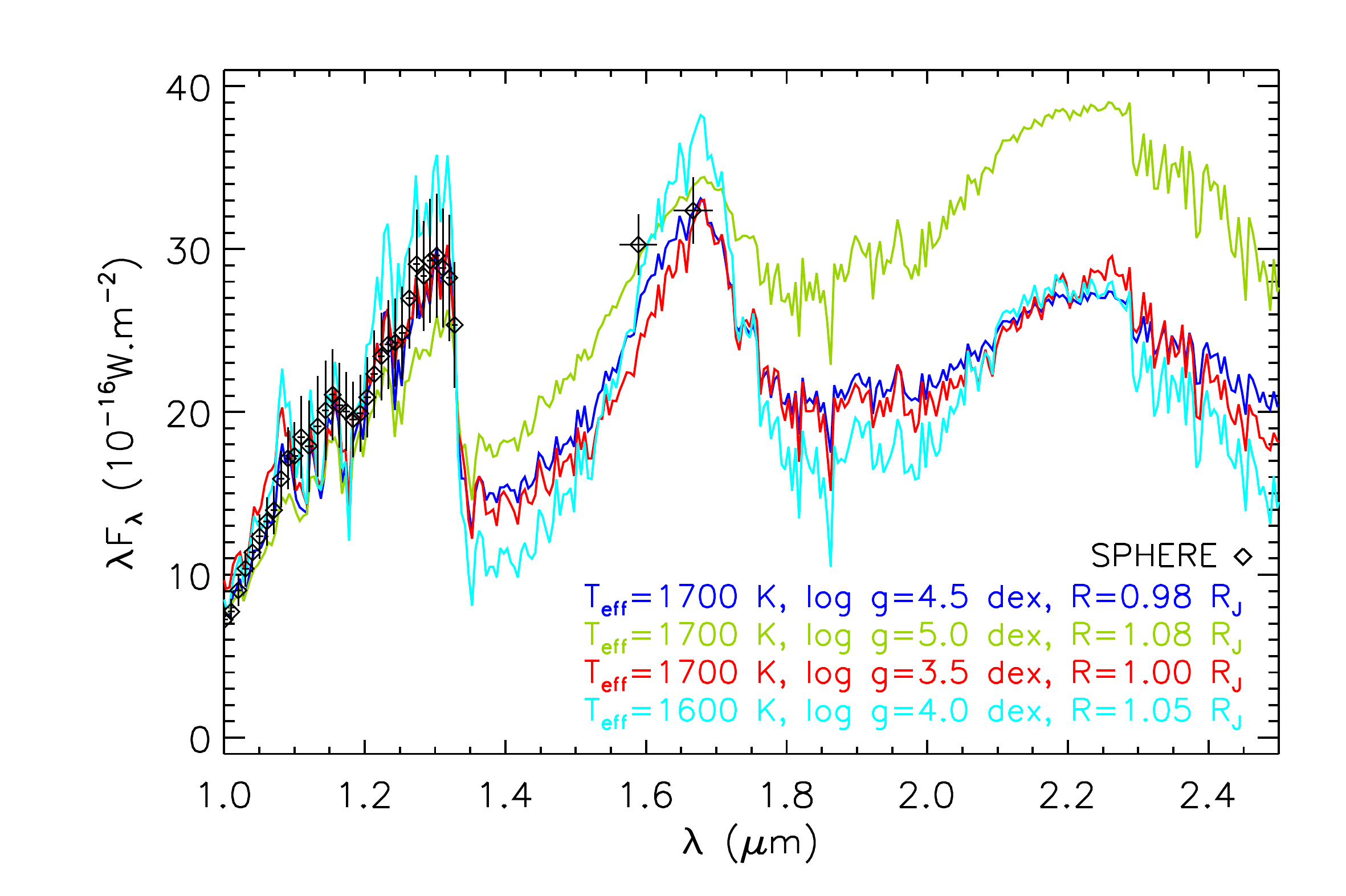}
\caption{\rev{Spectral energy distribution of HD\,72946B (black). The four best-match BT-Settl spectra are shown for comparison (colors).}}
\label{fig:sed}
\end{figure}

We show in Fig.~\ref{fig:sed} the resulting SED of HD\,72946B. We \rev{performed} a grid search for best-fit models in the BT-Settl spectral library \citep{Allard2011}. The characteristics of the grid \rev{are} $T_{\rm{eff}}$\,=\,700--2500~K by steps of 100~K, log\,$g$\,=\,3.5--5.5~dex by steps of 0.5~dex, and [Fe/H]\,=\,0.0~dex. We \rev{allowed} the radius to vary and kept solutions with radii in the range 0.7-1.1~$R_{\rm{J}}$. We show the four best-match model spectra in Fig.~\ref{fig:sed}. An effective temperature of $\sim$1700~K provides a good match to the data, which is in the range expected from evolutionary models for an age of $\sim$1--3~Gyr given the dynamical mass. \rev{It also agrees with a spectral type of L5 from the relation for field dwarfs in \citet{Filippazzo2015} (left panel of their Fig. 15)}.

\section{Discussion}
\label{sec:compa_models}

\begin{figure}[t]
\centering
\includegraphics[width=.4\textwidth,trim = 8mm 3mm 5mm 8mm, clip]{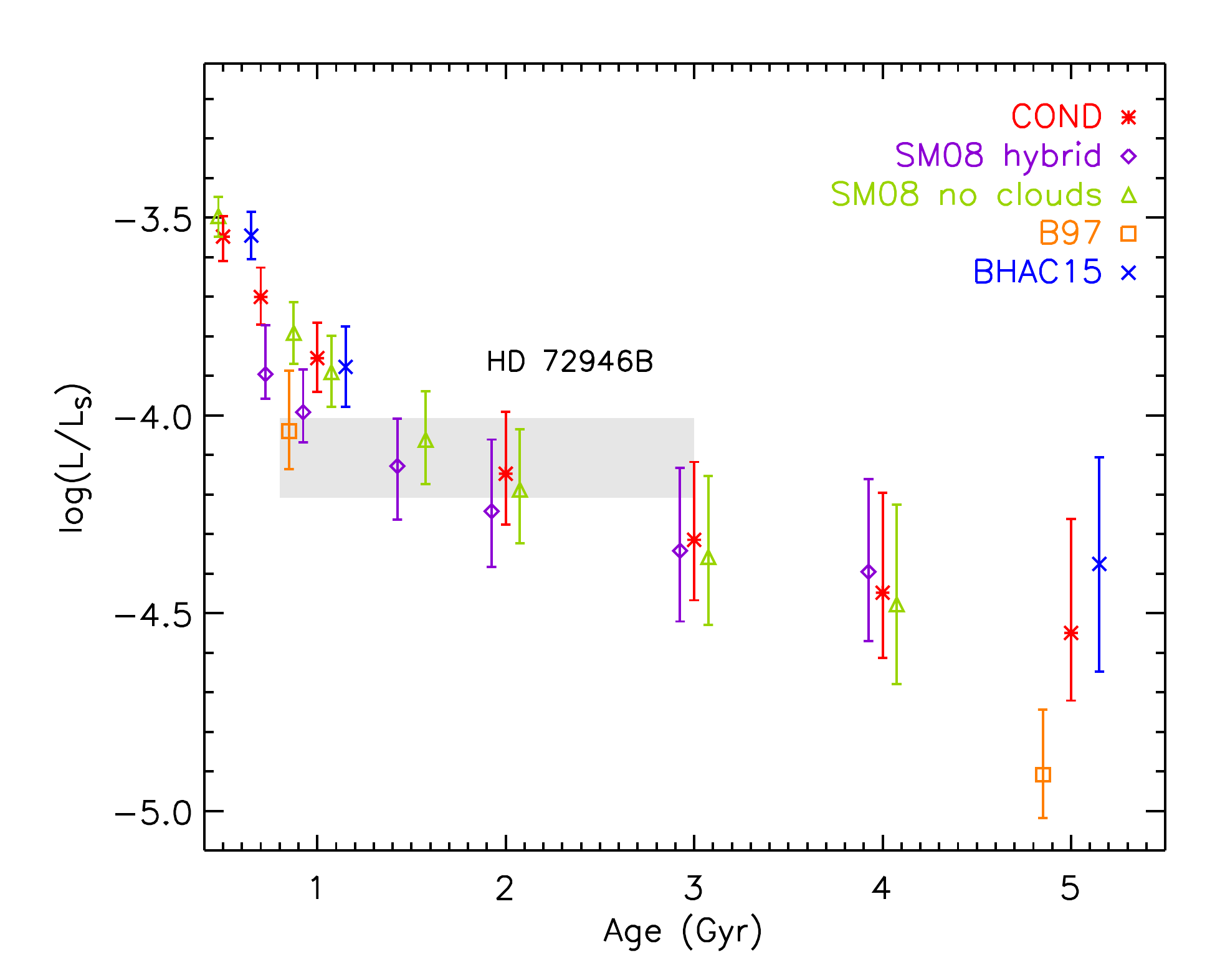}
\caption{\rev{Bolometric luminosity vs. age of HD\,72946B (gray area) compared to evolutionary tracks from the models COND \citep{Baraffe2003}, \citet{Saumon2008} (for two treatments of the clouds), \citet{Burrows1997}, and \citet{Baraffe2015} assuming the mass range for the companion from the orbital fit (data points). Small horizontal offsets are applied to all models except for COND for clarity.}}
\label{fig:lbol_age}
\end{figure}

HD\,72946B joins the short list of benchmark brown dwarf companions to stars with RV and imaging measurements: HR\,7672B \citep{Liu2002, Crepp2012}, HD\,19467B \citep{Crepp2014}, HD\,4747B \citep{Sahlmann2011, Crepp2016, Peretti2018}, GJ\,758B \citep{Thalmann2009, Bowler2018}, HD\,4113C \citep{Cheetham2018a}, and GJ\,229B \citep{Nakajima1995, Brandt2019c}. HD\,72946B stands out among these objects because a complete orbit is covered by RV and it has the smallest physical separation to the star, $\sim$6.4-6.5~au. This is slightly outside the ice line for a Sun-like star.

To evaluate a possible formation mechanism for HD\,72946B, we \rev{compared} its mass (or mass ratio to the star) and separation to model objects formed by fragmentation of a collapsing cloud in \citet{Bate2009} (Fig. 21) or by disk gravitational instabilities in \citet{Forgan2013} and \citet{Vigan2017} \rev{(left panel of Fig. 8 in the latter paper)}. \rev{The semi-major axes of most of the model objects with mass ratios similar to HD\,72946B formed in the former process are in the range 20--5000~au. The semi-major axes of most of the model objects with masses similar to HD\,72946B formed in the latter process are in the range $\sim$10--50~au. This means that the semi-major axis of HD\,72946B is smaller than those of model objects from both formation mechanisms, and we cannot exclude any of them}.

Figure~\ref{fig:lbol_age} shows the estimated bolometric luminosity and age of HD\,72946B with the predictions from the models COND \citep{Baraffe2003}, \citet{Saumon2008} (for two treatments of the clouds, hybrid and no clouds), \citet{Burrows1997}, and \citet{Baraffe2015} assuming the 95.4\% confidence interval for the companion mass from the orbital fit (72.4$\pm$3.2~$M_{\rm{J}}$). We \rev{estimate} the bolometric luminosity to be log($L$/$L_{\sun}$)\,=\,-4.11$\pm$0.10~dex using the magnitude-bolometric luminosity relation in \citet{Filippazzo2015} for field dwarfs and the $J_s$ magnitude computed from the IFS spectrum (15.44$\pm$0.13~mag) with a correction of 0.05~dex between the $J_s$ and $J$ bands estimated using SpeX spectra of the three best-fit template dwarfs. The $J$-$H3$ color of the companion (1.08$\pm$0.08~mag) is consistent with expectations from mid-L field dwarfs \citep{Cheetham2019} and is \rev{closer to the color predicted given the mass and \rev{age} of the companion by the DUSTY model \citep[cloudy atmosphere,][$J$-$H3$>1~mag]{Chabrier2000} than to the color predicted by the COND model (cloudless atmosphere, $J$-$H3$<0.8~mag)\footnote{Tables available at \url{http://perso.ens-lyon.fr/france.allard/}.}. This suggests a cloudy atmosphere}. For ages younger than 800~Myr, HD\,72946B is fainter than the predictions of all evolutionary models. At 1~Gyr, the companion properties are best reproduced by the hybrid cloud model of \citet{Saumon2008} and \citet{Burrows1997}. At 2~Gyr, the best-match models are COND and the cloudless model of \citet{Saumon2008}, and \citet{Baraffe2015}. At 3~Gyr, the models of \citet{Baraffe2015} account better for the companion properties. Observations to better constrain the stellar age with gyrochronology may allow a better distinction between the models.

The characterization of HD\,72946B clearly illustrates the improvements in the high-contrast imaging instrumentation toward bridging the gap in separation to the star with RV and astrometry. The combination of these data provides stronger constraints on the properties of substellar companions than can be reached with one technique alone. This allows testing their mass-luminosity models. The SPHERE data are sensitive to low-mass brown dwarfs down to $\sim$30~$M_{\rm{J}}$ at separations as close as 0.2$''$ (Appendix~\ref{sec:detlims_sphere}). The next generation of high-contrast imaging instruments on extremely large telescopes will enable extending analyses like this to the bulk of substellar companions that are detected with RV at closer separations and at lower masses down to the planetary regime and building empirical mass-luminosity relations for exoplanets. \rev{The future release of the \textit{Gaia} epoch astrometry will permit more accurate measurements of proper motion anomalies. This will improve dynamical mass estimates and provide new targets for this purpose.}

\begin{acknowledgements}
     The authors thank the referee for a constructive report that helped to clarify the manuscript. The authors thank the ESO Paranal Staff for support in conducting the observations. N.\,G. and J.\,O. acknowledge financial support from the ICM (Iniciativa Cient\'ifica Milenio) via the N\'ucleo Milenio de Formaci\'on Planetaria grant, from the Universidad de Valpara\'iso, and from Fondecyt (grant 1180395). N.~G. acknowledges grant support from project CONICYT-PFCHA/Doctorado Nacional/2017 folio 21170650. We acknowledge financial support from the Programme National de Plan\'etologie (PNP), the Programme National de Physique Stellaire (PNPS) of CNRS-INSU in France, and the Agence Nationale de la Recherche (GIPSE project; grant ANR-14-CE33-0018). T.H. acknowledges support from the European Research Council under the Horizon 2020 Framework Program via the ERC Advanced Grant Origins 83 24 28. This work made use of recipes from the IDL libraries: IDL Astronomy Users's Library \citep{Landsman1993}, Coyote (\url{http://www.idlcoyote.com/index.html}), EXOFAST \citep{Eastman2013}, JBIU (\url{http://www.simulated-galaxies.ua.edu/jbiu/}), and the s3drs package (\url{http://www.heliodocs.com/xdoc/index.html}). It also made use of the Python packages: NumPy \citep{Oliphant2006}, emcee \citep{ForemanMackey2013}, corner \citep{Foreman2016}, Matplotlib \citep{Hunter2007}, Astropy \citep{AstropyCollaboration2013, AstropyCollaboration2018}, and dateutil (\url{http://dateutil.readthedocs.io/}). This publication makes use of VOSA, developed under the Spanish Virtual Observatory project supported by the Spanish MINECO through grant AyA2017-84089.
VOSA has been partially updated by using funding from the European Union's Horizon 2020 Research and Innovation Programme, under Grant Agreement nº 776403 (EXOPLANETS-A). This research has benefitted from the SpeX Prism Spectral Libraries, maintained by Adam Burgasser at \url{http://pono. ucsd.edu/~adam/browndwarfs/spexprism}. This research made use of the SIMBAD database and the VizieR Catalogue access tool, both operated at the CDS, Strasbourg, France. The original descriptions of the SIMBAD and VizieR services were published in \citet{Wenger2000} and \citet{Ochsenbein2000}. This research has made use of NASA's Astrophysics Data System Bibliographic Services.
     
\end{acknowledgements}

%
   \bibliographystyle{aa} 
   \bibliography{biblio} 
%

\begin{appendix}

\section{Stellar multiplicity}
\label{sec:multiplicity}

\citet{Bouchy2016} noted that HD\,72946 is part of a multiple system. HD\,72945 is a comoving F8 star located at a projected separation of 230 au. \citet{Duquennoy1991} \rev{identified} it as a spectroscopic binary SB1 with a period of 14.3~d.

The Gaia parallaxes of HD\,72945 and HD\,72946 differ at the 4.2$\sigma$ level, indicating a distance difference along the line of sight of about 0.41$\pm$0.10~pc. The similar proper motion and systematic velocities of the two stars argue in favor of a physical association \citep[see also][]{Oh2017}. We speculate that the separation along the line of sight could be significantly larger than the one projected on the plane of the sky. Alternatively, the parallaxes of one or both components can be altered above the formal errors by the presence of the companions.

In addition, \citet{Dommanget2002} \rev{reported} three stellar companions with angular separations of 93$''$, 117$''$, and 122$''$.

Using the \textit{Gaia} parallaxes, we find that the three components identified by \citet{Dommanget2002} ($\pi$\,$<$\,12~mas) do not form a system with HD\,72945 and HD\,72946 ($\pi$\,$\sim$\,38--39~mas). Instead, we note in the \textit{Gaia} catalog a star (2MASS ID 08354678+0635294) at $\sim$130$''$ ($\sim$3400~au) from HD\,72946 with a parallax of 38.8196$\pm$0.0584~mas (distance along the line of sight $\sim$0.11$\pm$0.05~pc) and similar proper motion, but without a measured RV. Therefore we argue that the system is formed by four stellar and one substellar components. For the orbital analysis, we \rev{assumed} that the acceleration seen in the proper motion of HD\,72946 is entirely due to the substellar companion HD\,72946B.

The other components may provide additional constraints on the age of the system. In particular, we \rev{derive} for HD\,72945 an age and a mass using the PARAM web interface, the $T_{\rm{eff}}$ derived by \citet{Casagrande2011} from Str\"omgren photometry, \rev{the \textit{Gaia} parallax, the $V$ magnitude from \textsc{Hipparcos} (5.92$\pm$0.01~mag), and} the metallicity measured by \citet{Bouchy2016} for HD\,72946 \citep[0.01-dex difference only with the metallicity of HD\,72945 measured by][]{Casagrande2011}. We \rev{find} an age of 1.584$\pm$0.952~Gyr and a mass of 1.245$\pm$0.030~$M_{\sun}$. For this computation, we \rev{assumed} that the spectroscopic companion of HD\,72945 does not contribute significantly to the integrated photometry. This should be the case if its mass is not much higher than the expected minimum mass from the RV orbit. Using the SB9 orbit \citep{Pourbaix2004} and the isochronal mass above, we \rev{derive} a minimum mass for the secondary of 0.34~$M_{\sun}$. In any case, a significant contribution to the photometry would shift the measured isochronal age toward higher values than expected.
We report archival GPI data of HD\,72945 in Appendix~\ref{sec:gpi}.

\section{Comparison of extracted spectrophotometry}
\label{sec:compa_algos}

\begin{figure}[h]
\centering
\includegraphics[width=.44\textwidth]{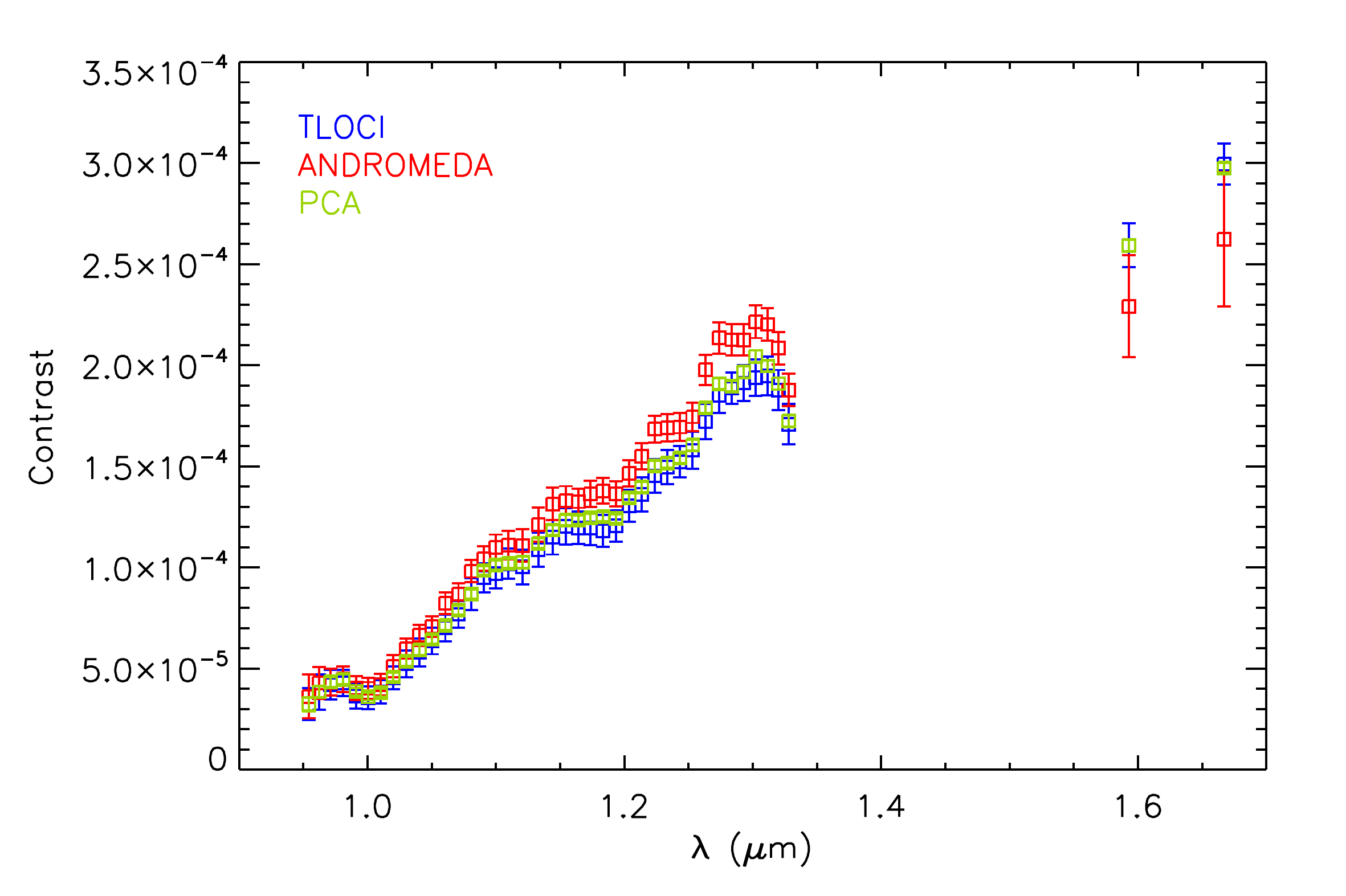}
\caption{Comparison of the TLOCI, ANDROMEDA, and PCA spectrophotometry. The measurement uncertainties are shown at 3~$\sigma$ for all algorithms.}
\label{fig:tloci_andro_spectra}
\end{figure}

We show in Fig.~\ref{fig:tloci_andro_spectra} the comparison of the spectrophotometry extracted with the ANgular DiffeRential Optimal Method Exoplanet Detection Algorithm \citep[ANDROMEDA,][]{Mugnier2009, Cantalloube2015} and with the Template Locally Optimized Combination of Images \citep[TLOCI,][]{Marois2014} and Principal Component Analysis \citep[PCA,][]{Soummer2012} algorithms provided by the SpeCal pipeline \citep{Galicher2018}. For the TLOCI extraction, we \rev{used} the fitting of a model planet image, whereas for PCA we \rev{employed} the negative synthetic planet injection. The fitting uncertainties are given at 3~$\sigma$. We note the good agreement between the TLOCI and PCA results within the TLOCI uncertainties. The IFS spectra between TLOCI and ANDROMEDA did not formerly agree for wavelengths longer than $\sim$1.2~$\muup$m, with the ANDROMEDA spectrum showing a steeper slope than the TLOCI spectrum. The IRDIS photometry for ANDROMEDA looks fainter than the TLOCI photometry by $\sim$15\%, although the uncertainties are large. This results in an IRDIFS spectrum that is redder for TLOCI than for ANDROMEDA.

We tested both ANDROMEDA and TLOCI SED for the atmospheric fitting. We experienced convergence problems when fitting the ANDROMEDA SED, and we chose the TLOCI SED for the analysis shown in this paper.
We \rev{did} not notice any significant discrepancies in the extracted astrometry, but we \rev{chose} to use the TLOCI astrometry for consistency.

\section{Corner plot of the orbital fit}
\label{sec:cornerplotorbit}

We provide here the corner plot of the orbital parameters derived in Sect.~\ref{sec:orbit}.

\begin{figure*}[t]
\centering
\includegraphics[width=.95\textwidth]{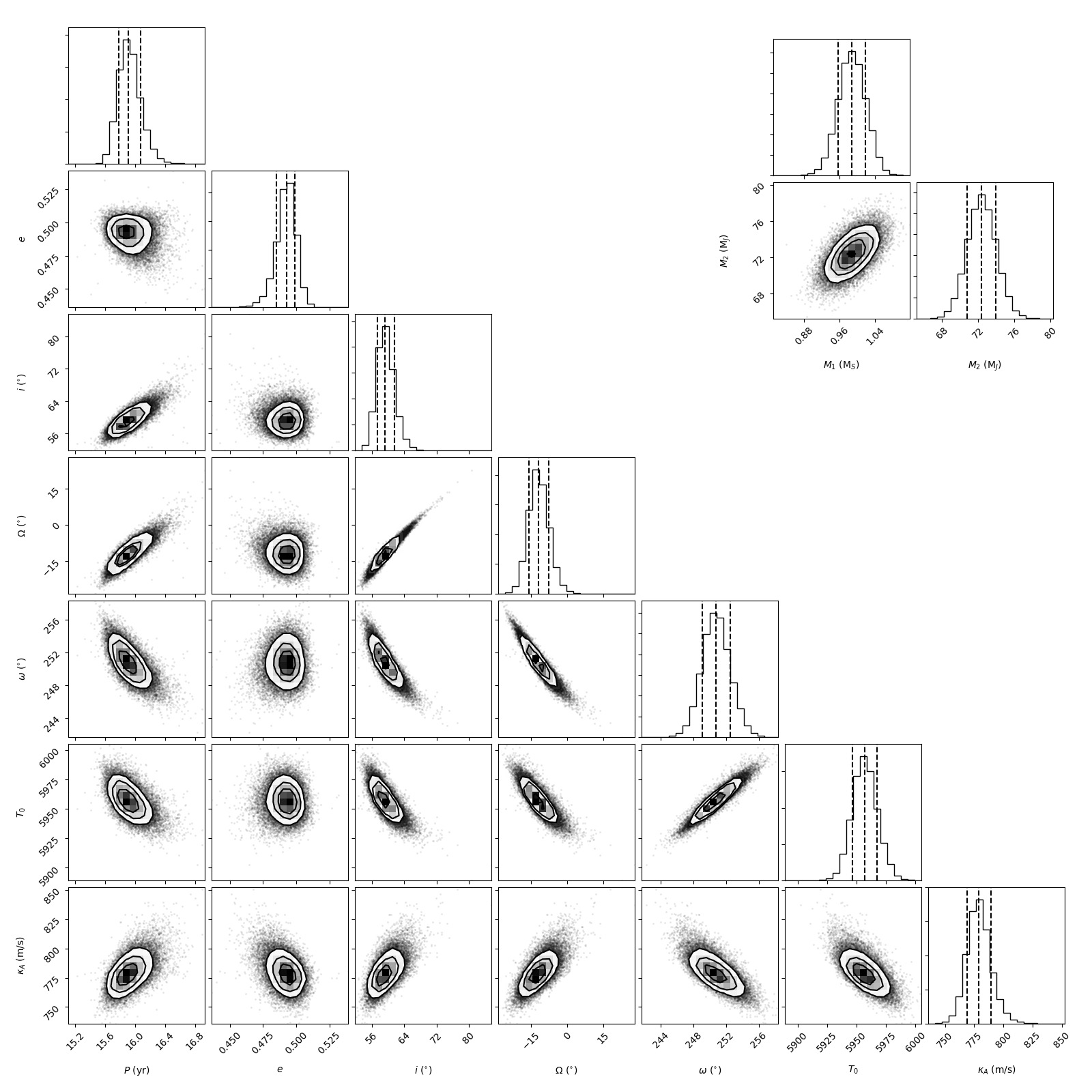}
\caption{MCMC posteriors of the orbital parameters (\textit{left}) and of the masses of HD\,72946 A and B (\textit{top right}). The diagrams displayed on the diagonals represent the 1D histogram distributions for the parameters. The off-diagonal diagrams show the correlations. In the histograms, the dashed vertical lines indicate the 16\%, 50\%, and 84\% quantiles.}
\label{fig:cornerplot}
\end{figure*}

\section{Color-magnitude diagram}
\label{sec:cmd}

\begin{figure}[h]
\centering
\includegraphics[width=.48\textwidth, trim = 6mm 3mm 5mm 5mm, clip]{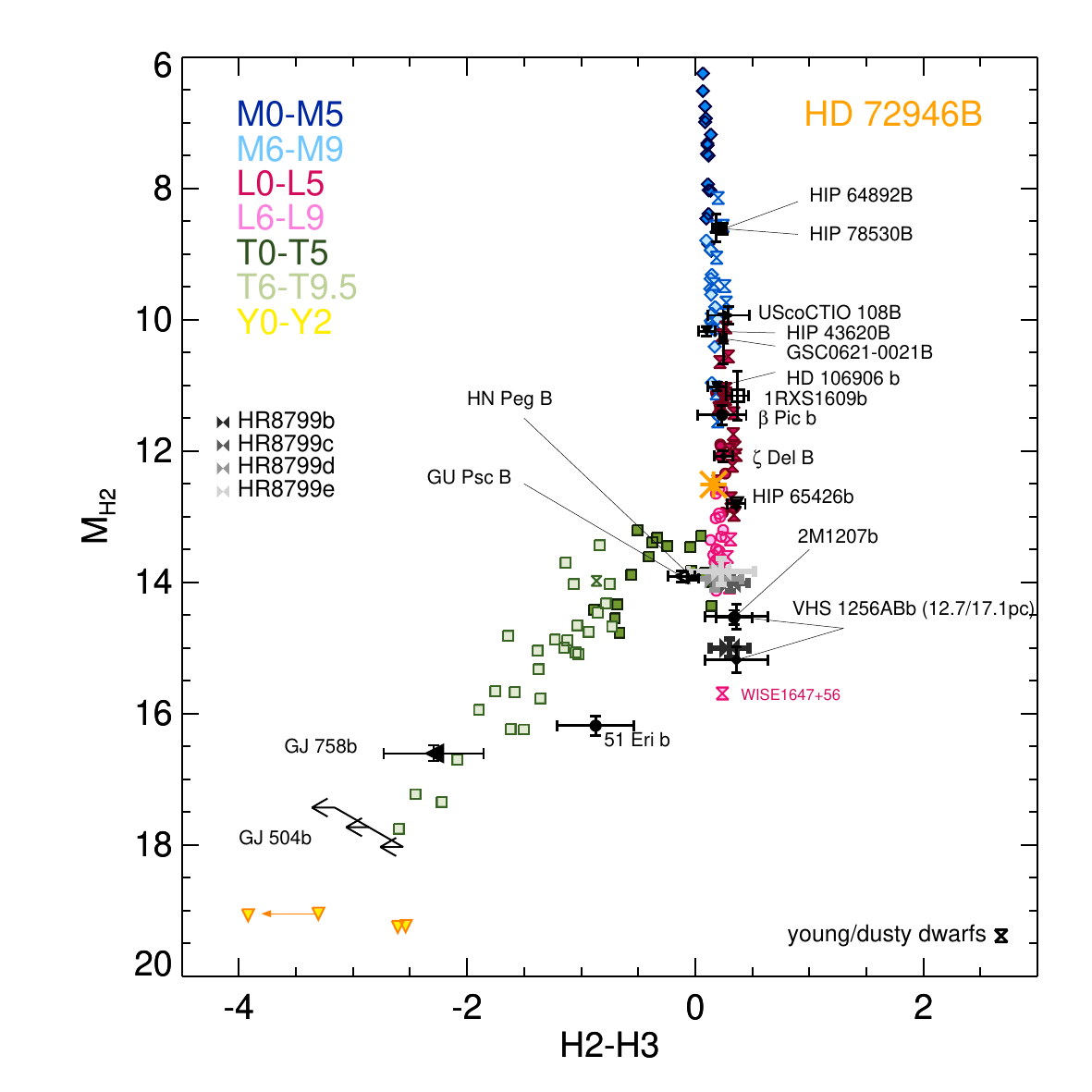}
\caption{\rev{Color-magnitude diagram of HD\,72946B (orange) using the IRDIS photometry. Template dwarfs (colored points) and a few young low-mass companions (black labels) are also shown for comparison.}}
\label{fig:cmd}
\end{figure}

To build the diagram in Fig.~\ref{fig:cmd}, we \rev{used} spectra of M, L, and T dwarfs from the SpeX-Prism library \citep{2014ASInC..11....7B} and from \citet{Leggett2000} and \citet{ASchneider2015} to generate synthetic photometry in the SPHERE filter passbands. The zero-points \rev{were} computed using a flux-calibrated spectrum of Vega \citep{Hayes1985, Mountain1985}. We also \rev{considered} the spectra of young and/or dusty free-floating objects from \citet{Liu2013}, \citet{Mace2013}, \citet{Gizis2015}, and of young companions \citep{2011ApJ...729..139W, Gauza2015, Stone2016, DeRosa2014, Lachapelle2015, Bailey2014, Rajan2017, Bonnefoy2014b, Patience2010, Lafreniere2010, Chauvin2017a, Delorme2017a, Cheetham2018b}. The colors and absolute fluxes of the benchmark companions and isolated T-type objects were generated from the distance and spectra of these objects in Appendix B in \citet{Bonnefoy2018}. To conclude, we \rev{used} the spectra of Y dwarfs published in \citet{ASchneider2015}, \citet{Warren2007}, \citet{Delorme2008}, \citet{Burningham2008}, \citet{Lucas2010}, \citet{Kirkpatrick2012}, and \citet{Mace2013} to extend the diagrams in the late-T and early-Y dwarf domain. We \rev{used} the distances of the field dwarfs reported in \citet{Kirkpatrick2000}, \citet{Faherty2012}, \citet{Dupuy2013}, \citet{Tinney2014}, \citet{Beichman2014}, and \citet{Luhman2016}. We \rev{considered} those reported in \citet{Kirkpatrick2011}, \citet{Faherty2012}, \citet{ZapateroOsorio2014}, and \citet{Liu2016} for the dusty dwarfs. The companion distances were taken from \citet{vanLeeuwen2007} and \citet{Ducourant2014}.

\section{SPHERE detection limits}
\label{sec:detlims_sphere}

\begin{figure}[t]
\centering
\includegraphics[width=.4\textwidth, trim = 8mm 4mm 5mm 8mm, clip]{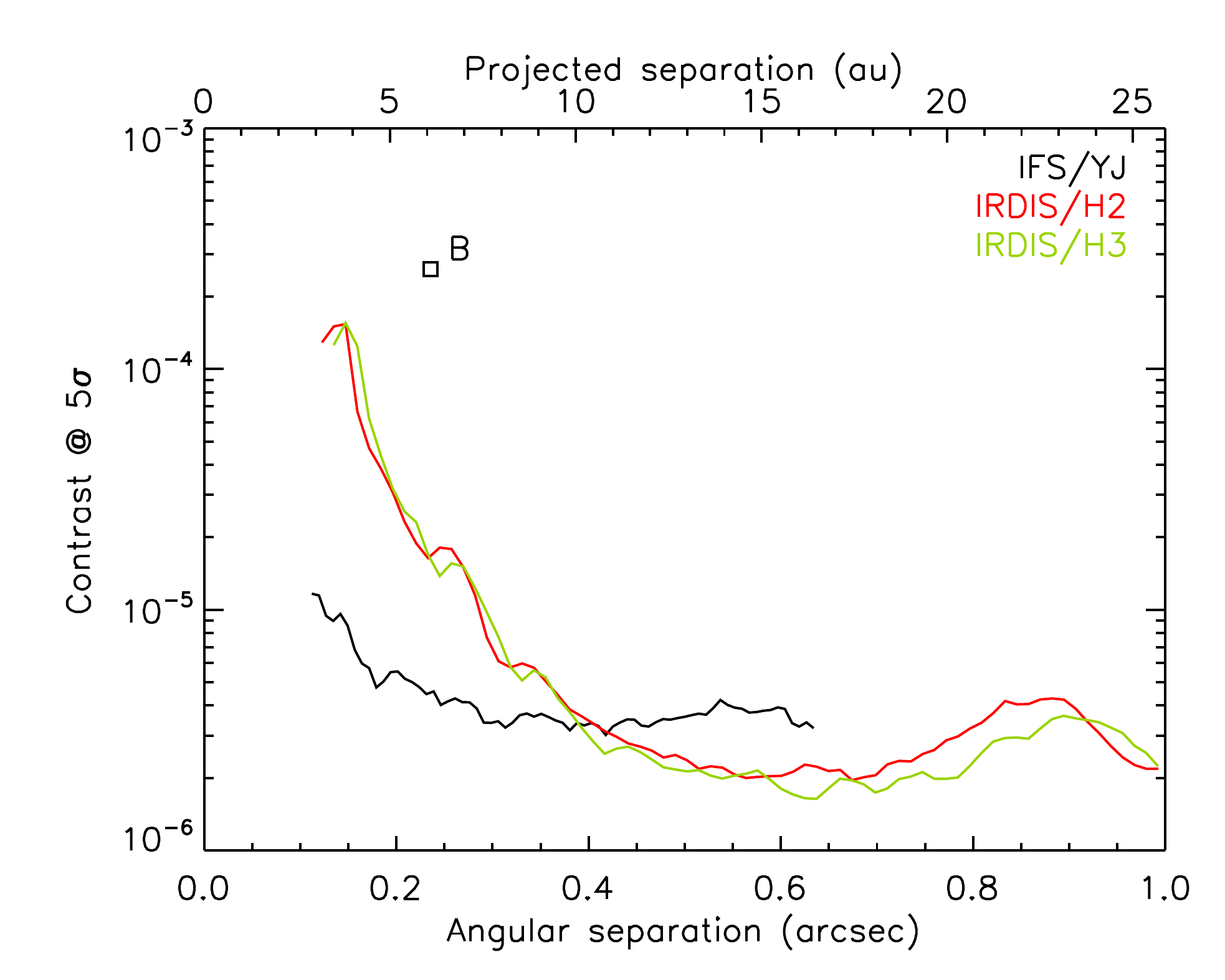}
\includegraphics[width=.4\textwidth, trim = 8mm 4mm 5mm 0mm, clip]{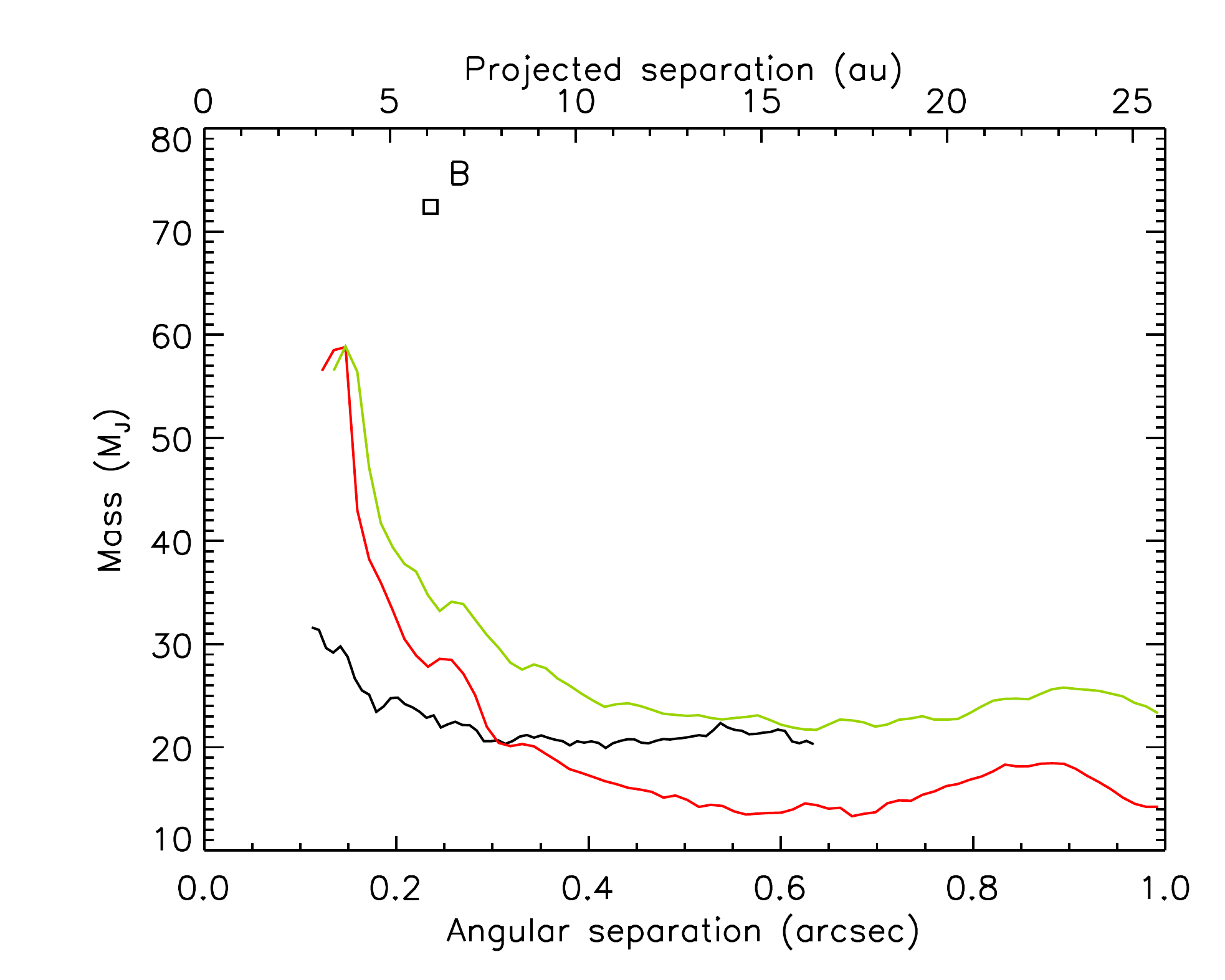}
\caption{\rev{SPHERE 5$\sigma$ detection limits in contrast with respect to the star (\textit{top}) and the planet mass (\textit{bottom}). We also indicate the location of HD\,72946B for comparison.}}
\label{fig:detlims}
\end{figure}

Figure~\ref{fig:detlims} shows the SPHERE detection limits in contrast to the star and the planet mass obtained with ANDROMEDA. For the IFS detection limits, we \rev{assumed} a T5 dwarf template spectrum (Cantalloube et al., in preparation) retrieved from the SpeX library. The detection limits account for the coronagraphic transmission \citep{Boccaletti2018} and the small sample statistics correction \citep{Mawet2014}. The contrast to planet mass conversion was derived assuming the ``hot-start'' evolutionary and atmospheric models of \citet{Baraffe2003, Baraffe2015} and an age of 2~Gyr for the system \citep[table from][]{Vigan2015}. The H3 curve is sensitive to more massive objects than the H2 curve because the probed mass regime corresponds to cold objects with strong methane absorption features and the H3 filter matches a strong methane band. We excluded additional brown dwarf companions more massive than $\sim$20~$M_{\rm{J}}$ at separations beyond 8~au.

\section{GPI archival data of HD\,72945}
\label{sec:gpi}

\begin{figure}[t]
\centering
\includegraphics[width=.4\textwidth, trim = 8mm 4mm 5mm 8mm, clip]{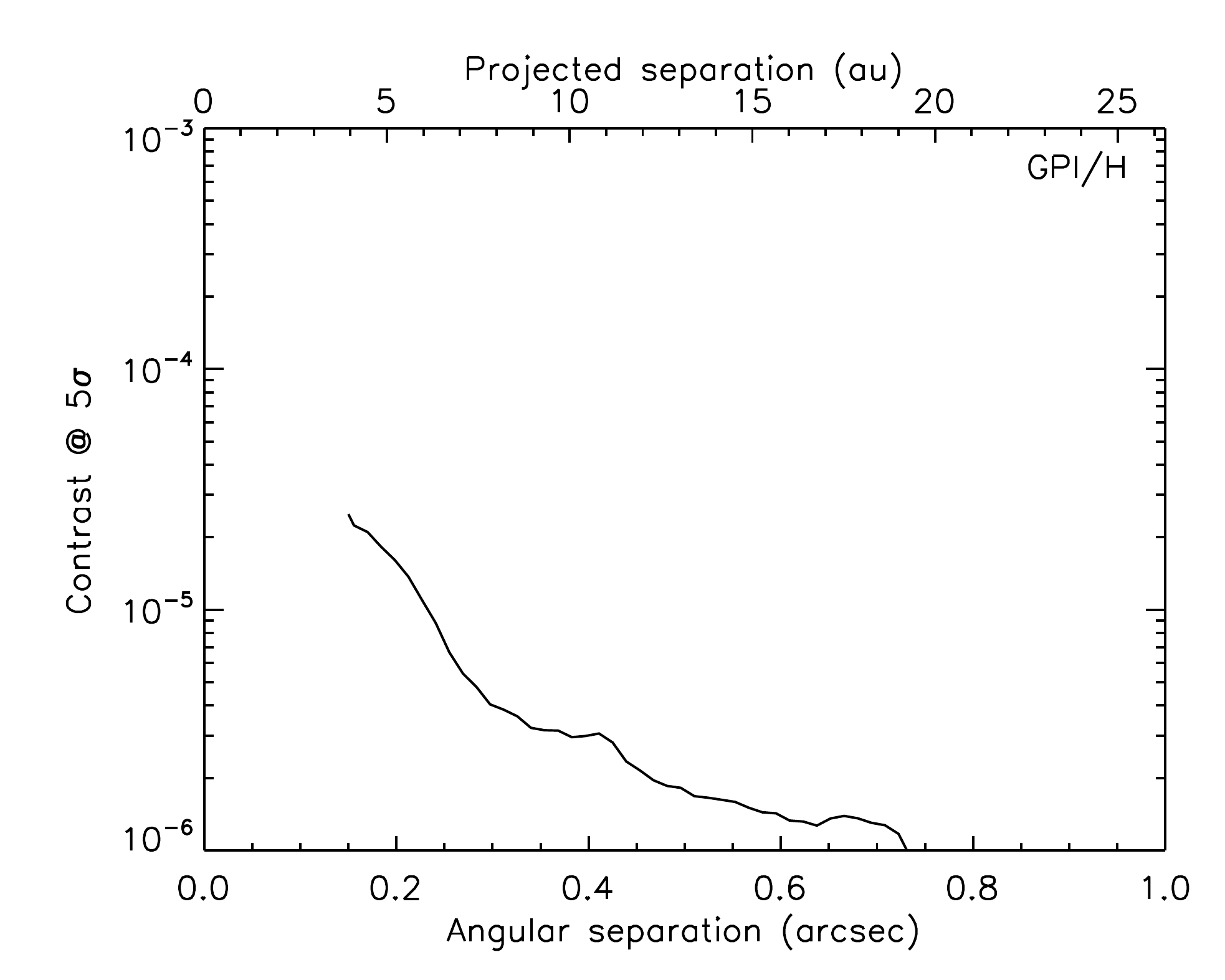}
\includegraphics[width=.4\textwidth, trim = 8mm 4mm 5mm 0mm, clip]{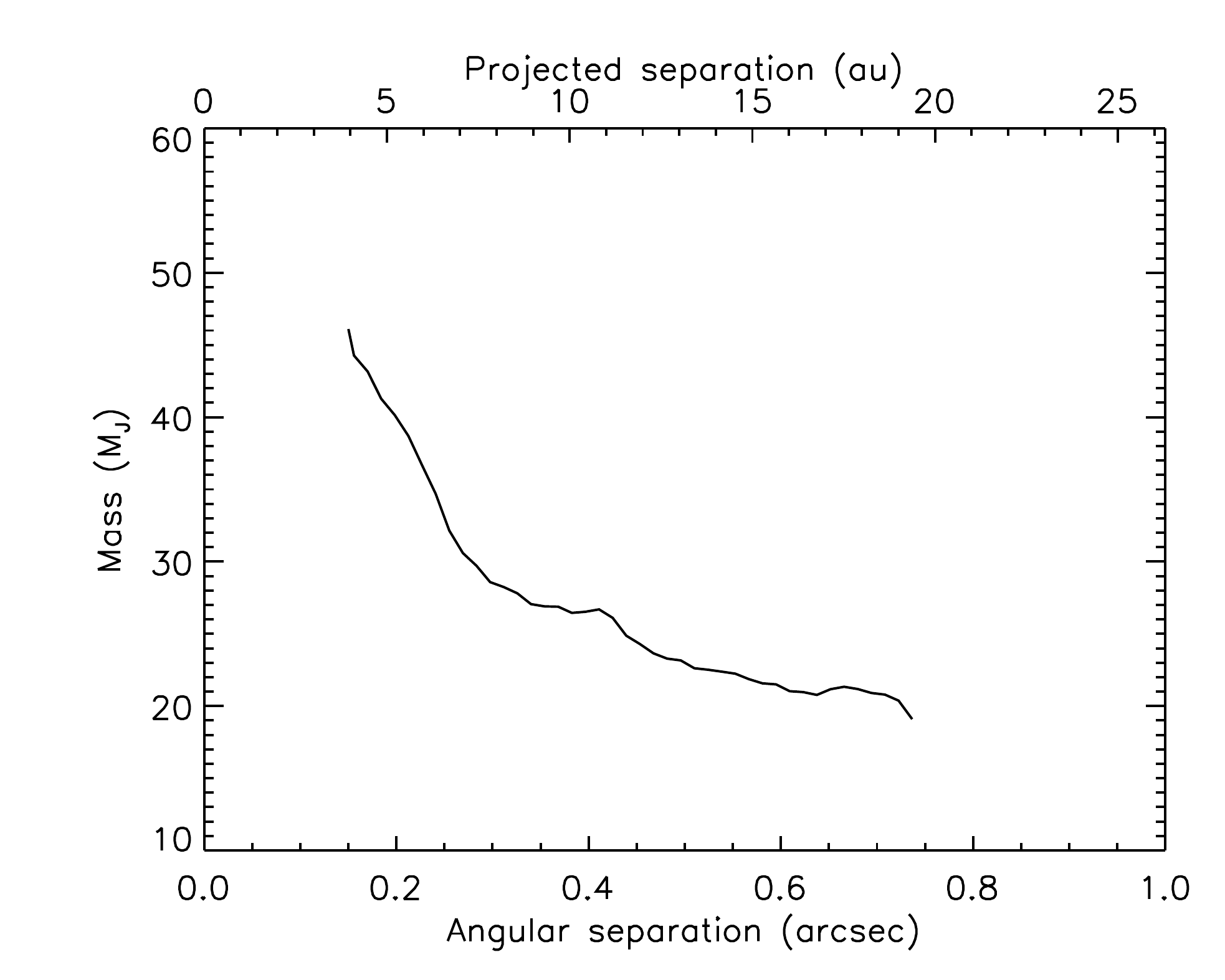}
\caption{GPI 5$\sigma$ detection limits of HD\,72945 in contrast with respect to the star (\textit{top}) and the planet mass (\textit{bottom}).}
\label{fig:detlims_gpi}
\end{figure}

The stellar SB companion HD\,72945 was observed with the Gemini Planet Imager \citep[GPI,][]{Macintosh2014} on 2015 April 4 UT in the $H$ band. The data were presented in the first statistical analysis of the GPIES survey \citep[][target name: HR~3395]{Nielsen2019}. The target was observed for an integration time of 32.8~min, which amounts to a field rotation of 19.5$^{\circ}$.

We retrieved the data from the Gemini archive and reduced them with the GPI data reduction pipeline v1.4.0 \citep{Perrin2014, Perrin2016}, which applies an automatic correction for the North offset of $-$1.00$\pm$0.03$^{\circ}$ measured by \citet{Konopacky2014}. Then, we post-processed them using ANDROMEDA. No point source is detected above 5$\sigma$. We show in Fig.~\ref{fig:detlims_gpi} the detection limits obtained for a T5 dwarf template spectrum. We \rev{assumed} an age of 2~Gyr, a distance of 26.3~pc from the \textit{Gaia} DR2 parallax, and the models of \citet{Baraffe2003, Baraffe2015}. For the stellar magnitude, we \rev{used} the 2MASS value, although we note that it is affected by saturation. We \rev{included} the small sample statistics correction. We cut the curves to separation larger than 0.15$''$ because we were unable to find GPI coronagraphic transmission curves. 

\end{appendix}

\end{document}